# Tunable optical absorption and interactions in graphene via oxygen plasma

**Iman Santoso**[1,2,4,6], **Ram Sevak Singh**[2], **Pranjal Kumar Gogoi**[1,2,3], **Teguh Citra Asmara**[1,2,3], **Dacheng Wei**[2], **Wei Chen**[2,4,5], **Andrew T.S. Wee**[2,4], **Vitor M. Pereira**[2,4,#], **Andrivo Rusydi**[1,2,3,*]

[1]*NUSNNI-NanoCore, National University of Singapore, Singapore 117576*

[2]*Department of Physics, National University of Singapore, Singapore 117542*

[3]*Singapore Synchrotron Light Source, National University of Singapore, 5 Research Link, Singapore 117603, Singapore*

[4]*Graphene Research Centre, Faculty of Science, National University of Singapore, Singapore 117546*

[5]*Department of Chemistry, National University of Singapore, Singapore 117543*

[6]*Jurusan Fisika,FMIPA, Universitas Gadjah Mada, BLS 21 Yogyakarta 55281, Indonesia*

[*]phyandri@nus.edu.sg

[#]vpereira@nus.edu.sg

**Abstract**

We report significant changes of optical conductivity ($\sigma_1$) in single layer graphene induced by mild oxygen plasma exposure, and explore the interplay between carrier doping, disorder, and many-body interactions from their signatures in the absorption spectrum. The first distinctive effect is the reduction of the excitonic binding energy that can be extracted from the renormalized saddle point resonance at 4.64 eV. Secondly, $\sigma_1$ is nearly completely suppressed ($\sigma_1 \ll \sigma_0$) below an exposure-dependent threshold in the near infrared range. The clear step-like suppression follows the Pauli blocking behaviour expected for doped monolayer graphene. The nearly zero residual conductivity below $\omega \sim 2E_F$ can be interpreted as arising from the weakening of the electronic self-energy. Our data shows that mild oxygen exposure can be used to controllably dope graphene without introducing the strong physical and chemical changes that are common in other approaches to oxidized graphene, allowing a controllable manipulation of the optical properties of graphene.

## I.  Introduction

In the absence of disorder, the optical conductivity of graphene –carbon atoms arranged in a two dimensional hexagonal lattice – displays many remarkable optical properties, including a broadband universal optical conductivity ($\sigma_0 = \pi e^2/2h$) in the infrared-to-visible range [1-4]. In the visible–ultraviolet range the interplay between electron-electron (e-e) and electron-hole (e-h) interactions yields unique excitonic effects that renormalize and redshift the bare band structure saddle point resonance by ~0.6 eV. This is clearly seen in the real part of the optical conductivity ($\sigma_1$) of graphene [3-9], and is one of the clear instances where explicit consideration of many-body interactions is required for an accurate description of the electronic properties of graphene. Recently, controlled disorder, such as defects, impurities, vacancies, and adatoms, has been studied intensively and proposed as a means to tailor the transport and optical properties of graphene [10-15]. For example, the functionalization of graphene using K, Rb [16] and $NO_2$ [18] have shown the large doping effects indicated by strong suppression of the absorption below $2E_F$ ($E_F$ is Fermi energy). Here we explore the optical response of graphene to controlled disorder induced by oxygen plasma exposure.

Mild oxygen plasma exposure has been widely used to produce graphene oxide. This dry method has numerous advantages, namely: the oxidation can take place rapidly, and it does not strongly modify the transport properties of graphene. It is known that this method introduces structural defects and electronic disorder due to the attachment of oxygen to carbon atoms, and to the reduction in the overall $sp^2$ order [18]. The fact that the oxygen arrives to the sample surface by a process of diffusion, makes remote oxygen plasma treatment a clean way to control the amount of disorder in graphene [18]. However, plasma exposure can have very different outcomes in the optical and transport properties of graphene, depending on the intensity of irradiation that directly translates into the degree of disorder and amorphization of the resulting carbon lattice. Increased carrier densities, semiconducting transport behaviour, and photoluminescence are frequently seen [18,19]. However, the electronic structure and correlation effects in graphene with controlled disorder through the optical spectroscopy technique remain unexplored. Hence, it is of great interest to study directly the optical absorption spectrum of these systems, and analyse the interplay between carrier doping, disorder, and many-body interactions from their signatures in the optical absorption spectrum.

Here, we study the evolution of optical conductivity, $\sigma_1(\omega)$, of monolayer graphene in the frequency range of 0.5–5.3 eV under mild (low power, ~6W) exposure to oxygen plasma. Single layer graphene (SLG) was prepared on the surface of a copper foil via a low-pressure chemical vapour deposition (CVD) process, and then transferred to a $SiO_2$ (300nm)/Si substrate. Exposure of SLG to the oxygen plasma is done in successive steps. After each 2-second ($t_s$) exposure step Raman scattering and spectroscopic ellipsometry (SE) measurements are performed on the sample. The Raman traces are used to follow the level of disorder and amorphization in the system following each exposure step. The spectroscopy ellipsometry measurements are used to directly measure $\sigma_1(\omega)$ in the spectral range 0.5–5.3 eV. As $t_s$ increases two important effects emerge in the optical conductivity: the decrease of the resonant saddle-point peak intensity at 4.6 eV, concomitant with a significant reduction of the excitonic binding energy, and the dramatic suppression of $\sigma_1$ (close to zero) for photon energies below 1 eV.

## II.    Sample preparations and experimental methods

A low-pressure chemical vapour deposition (CVD) process was used to prepare SLG. A copper foil was placed in the center of a horizontal 2 inch quartz tube furnace, and then was annealed at 1000 °C in 10 sccm (150 mTorr ) $H_2$ for 30 min. After that, 30 sccm $CH_4$ was introduced into the furnace as the carbon source for SLG growth, while the $H_2$ was maintained at 10 sccm and the furnace temperature maintained at 1000 °C. The pressure was maintained at 350 mTorr in the growth. After 30 minutes the $CH_4$ was turned off and the furnace was quickly cooled down to room temperature in 10 sccm $H_2$. To transfer SLG to a $SiO_2$ (300 nm)/Si substrate a PMMA film was spin-coated on the copper foil with SLG. After that, the copper foil was etched in $FeCl_3$ aqueous solution for 12 hours, and then a PMMA film with SLG can be obtained on the solution surface. After cleaning the PMMA film in purified water for several times, it was transferred to a $SiO_2$ (300 nm)/Si substrate, and was heated at 180 °C for 2 min. To remove the PMMA the substrate was placed in acetone vapor for 15 min and was annealed at 300 °C in 10 sccm $H_2$ for 30 min. The SLG sample was then exposed to the mild oxygen plasma for three consecutive durations of 2 s, 4 s, and 6 s using a radio-frequency (rf) plasma system. The rf power was maintained at 6W and chamber pressure at 50 mTorr. The pristine SLG and oxygen plasma exposed SLG were investigated using Raman and spectroscopy ellipsometry (SE) measurements. Raman spectra were recorded using a 514 nm laser excitation wavelength (Renishaw Raman measurement system). A low laser power of ~ 2 mW is employed to minimize laser heating effects. All the measurements were performed at room temperature, in ambient, and on the same sample. To check the stability of the sample we placed the sample in a dry box for one day after SE measurements. After that, we repeated the SE measurements and the SE data showed no significant change, indicating that the sample is stable enough for measurements in an ambient environment.

Spectroscopic ellipsometry (SE) [34] measurements are performed on the graphene sample and substrate ($SiO_2$ (300 nm)/Si) using a SENTECH SE850 ellipsometer. This ellipsometer is equipped with three different light sources, i.e. Deep UV ((deuterium), UV/VIS source (Xe-lamp) and the NIR source (Halogen lamp of the FT-IR spectrometer), allowing us to measure from 0.5 eV to 6.3 eV. For our measurements we used additional micro-focus probes (~100 micron spot diameter) which work well in the range from 0.5 – 5.3 eV. SE allows the accurate measurement of ellipsometric parameters $\Psi$ and $\Delta$ defined from the relation $\tan(\Psi)\exp(i\Delta) = \frac{r_p}{r_s}$, where $r_p$ and $r_s$ are the complex reflection amplitudes for the *p* and *s* light polarizations. Thus, $\Psi$ represents the ratio between the amplitude of *p* and *s* polarized reflected light while $\Delta$ represents the phase difference between *p* and *s* polarized reflected light.

## III.    Results and discussion

In Figure 1 we show the Raman spectra for graphene on $SiO_2$ before (pristine) and after exposure to oxygen plasma for 2s, 4s, and 6s. The high quality of our pristine graphene is confirmed by the very sharp Raman G peak at 1593 cm$^{-1}$ with a full width at half maximum (FWHM) of 20 cm$^{-1}$, a 2D peak at 2694 cm$^{-1}$ with FWHM of 36 cm$^{-1}$, and no evidence of defect scattering (absent D peak). Furthermore, the 2D to G peak intensity ratio ($I_{2D}/I_G$) is around two, and the 2D peak possesses a symmetric single Lorentzian shape. This, together with the behaviour of the $\sigma_1$ (see below) indicates that our system consists of monolayer graphene [20]. Upon exposure to oxygen plasma, the appearance of the D and D' peaks at ~1318 cm$^{-1}$ and 1615 cm$^{-1}$, respectively, indicates pronounced inter- and intra-valley scattering [19-22]. After the first exposure, we observe a decrease in the ratio between the D and G peak intensities ($I_D/I_G$), which changes from 1.65 to 0.95 with growing $t_s$. The fact that the D and D' peaks are intense and broad for $t_s > 2s$ suggests relatively strong disorder. Furthermore, based on the three-stage model of structural disorder in graphitic materials [23], our oxidized samples are likely in stage two. In this stage the decrease of ($I_D/I_G$) is due to the decrease in the number of sp$^2$ ordered rings, and is related to the in-plane crystalline grain size ($L_a$) through the empirical formula: $I_D/I_G = C'(\lambda)L_a^2$, where $C'(\lambda)$ denotes a constant at the particular excitation wavelength ($\lambda$) used in Raman measurements. Here we use $C'(514\ nm) = 0.55\ nm^{-2}$ [18,23].

Spectroscopy Ellipsometry data for $\Psi$ and $\Delta$ are taken at multiple incident angles and at several spots on the sample. The data at different spots are identical in almost all cases which show sample homogeneity. The multiple incident angle data is required for global fitting of the data. SE measurement is generally preferable because it gives both the real and imaginary parts of the dielectric function directly, whereas other measurement techniques (such as direct reflectivity) require a Kramers-Kronig transformation. Moreover, in case of very thin films the change in phase of the incident light waves upon reflection is much more pronounced than the change in amplitude of the light of different polarizations. These two facts make SE an ideal method for analyzing systems like the very thin monolayer of graphene on a substrate. In Figure 2 we show the measured $\Psi$ and $\Delta$ values of samples with and without graphene on substrate at 70° incident angle. The spectra show a pronounced contrast due to the presence of graphene which is only ~3 Angstrom thick.

We show the experimental $\Psi$ and $\Delta$ of pristine graphene on the $SiO_2$ (300 nm)/Si substrate and after cumulative oxygen plasma exposure ($t_s$) in 2 seconds steps in Figs. 3 (a) and (b). Due to oxygen plasma exposure, $\Psi$ and $\Delta$ are slightly changed, especially in the energy range indicated in the inset of Figs. 3 (a) and (b). These facts show that the optical constants (e.g. dielectric constant and hence optical conductivity) changed and care should be taken when the fitting procedure is applied in these particular energy ranges. In contrast, within the error bars of the measurements, there is no significant change of the measured $\Psi$ and $\Delta$ of the $SiO_2$ (300

nm)/Si substrate after cumulative oxygen plasma exposure (see Figs. 3(c) and (d) as well as the insets).

To extract the optical constants, the system is modeled with Fresnel coefficients for an optical multilayered film system (see inset of Fig. 2 (a)). The model is composed of a silicon bulk (label 3), a 300 nm SiO$_2$ layer (label 2), a 0.335 nm graphene layer (label 1), and the air (label 0). The Fresnel coefficients of reflected light for each interface of two adjacent layers are given by [34-36]:

$$r_{01,p} = \frac{\sqrt{\varepsilon_1}\cos\theta_0 - \sqrt{\varepsilon_0}\cos\theta_1}{\sqrt{\varepsilon_1}\cos\theta_0 + \sqrt{\varepsilon_0}\cos\theta_1} \quad ; \quad r_{01,s} = \frac{\sqrt{\varepsilon_0}\cos\theta_0 - \sqrt{\varepsilon_1}\cos\theta_1}{\sqrt{\varepsilon_0}\cos\theta_0 + \sqrt{\varepsilon_1}\cos\theta_1} \quad , \tag{1}$$

where we have implicitly applied the Fresnel equation for the interface between medium 0 (air) and medium 1 (graphene layer) in deriving eq.(1). The *p* and *s* denote the *p* and *s* polarization of light, respectively. The $\varepsilon_0$ and $\varepsilon_1$ represent the dielectric constant of medium 0 and medium 1. $\theta_0$ and $\theta_1$ represent the incident angle and refracted angle at the interface between medium 0 and medium 1. These angles are related to each other by Snell's law. The Fresnel coefficients of reflected light for other interfaces (e.g. $r_{12}$ and $r_{23}$) can be obtained by substituting the index 0 and 1 in eq.(1) with the respective index under consideration. The total amplitude of the reflection of the system given in the inset of Fig. 2 (a) is [34-36]:

$$r_{0123,p(s)} = \frac{r_{01,p(s)} + r_{12,p(s)}\exp(-i2\beta_1) + [r_{01,p(s)}r_{12,p(s)} + \exp(-i2\beta_1)]r_{23,p(s)}\exp(-i2\beta_2)}{1 + r_{01,p(s)}r_{12,p(s)}\exp(-i2\beta_1) + [r_{12,p(s)} + r_{01,p(s)}\exp(-i2\beta_1)]r_{23,p(s)}\exp(-i2\beta_2)} \quad , \tag{2}$$

Where *p(s)* denotes *p(s)* polarized light; $\beta_1$ and $\beta_2$ represent the phase difference when light penetrates through the interface between medium 0 and 1, and between medium 2 and 3, respectively. $\beta_1$ and $\beta_2$ are given by $\beta_1 = \frac{2\pi d_1}{\lambda}[\varepsilon_1 - \varepsilon_0\sin^2\theta_0]^{1/2}$ and $\beta_2 = \frac{2\pi d_2}{\lambda}[\varepsilon_2 -$ $\varepsilon_0 \sin^2 \theta_0]^{1/2}$. $d_1$ and $d_2$ are the thickness of the layer 1 and layer 2 while $\lambda$ denotes the wavelength of the light source used in this experiment.

The dielectric constants of the $SiO_2$ and silicon bulk are extracted from a fit of the experimental $\Psi$ and $\Delta$ of the substrate with an optical model comprised of a 300 nm $SiO_2$ layer and silicon bulk (see the inset of Fig. 2 (a) with red arrow). In extracting these dielectric constants, we use many oscillators which can be described by Drude-Lorentz model [35,36]:

$$\varepsilon(\omega) = \varepsilon_\infty + \sum_k \frac{\omega_{p,k}^2}{\omega_{0,k}^2 - \omega^2 - i\gamma_k \omega} \ , \qquad (3)$$

where $\varepsilon_\infty$ denotes the high-frequency dielectric constant, which represents the contribution from all oscillators at very high frequencies compared to the frequency ranges under examination. The parameters $\omega_{p,k}$, $\omega_{o,k}$ and $\gamma_k$ are the plasma frequency, the transverse frequency (eigenfrequency) and the linewidth (scattering rate), respectively, of the k-th Lorentz oscillator. For the $SiO_2$ layer we have used one additional oscillator at high energy (~9 eV) in order to capture the first absorption peak in $SiO_2$ [37] and, hence, give the best match between the model and the experimental $\Psi$ and $\Delta$. Alternatively, one can use the Cauchy layer for describing the $SiO_2$ layer [38]. The obtained dielectric constants of the $SiO_2$ and silicon bulk used for extracting the dielectric constant of graphene based on the optical model are shown in the inset of Fig. 2 (a) (with a blue arrow). We assume that the graphene film is flat and isotropic [38-40]. Having extracted the complex $\varepsilon(\omega) = \varepsilon_1(\omega) + i\varepsilon_2(\omega)$, $\sigma_1(\omega)$ follows immediately from

$$\sigma_1 = \frac{\omega \varepsilon_2}{4\pi} \ . \qquad (4)$$

Figures 4 and 5 plot the detailed analysis of SE data for pristine graphene on $SiO_2$ (300 nm)/Si substrate and after 6 seconds of oxygen plasma exposure time ($t_s$), respectively. The shaded area in $\varepsilon_1$ and $\sigma_1$ (Figs. 5 (a) and (b)) indicates the range in which the model can still match the experimental $\Psi$ and $\Delta$ (within the error bars).

In the next discussion, we concentrate on the evolution of $\sigma_1$ as function of $t_s$ as shown in Figure 6. In addition to the Raman characterization, the monolayer character of our sample is further corroborated by the constancy and magnitude of $\sigma_1$ at low frequencies, which rather accurately coincides with the universal value $\sigma_0 = \pi e^2/2h$ expected for a graphene monolayer [31]. The two main effects of oxygen exposure are: (1) a significant reduction (more than 50 %) of the van Hove peak intensity at 4.64 eV in the ultraviolet (UV) range, accompanied by a large blue-shift of the peak position; (2) $\sigma_1$ is gradually suppressed to zero in a step-like fashion as $t_s$ increases in the near infrared (NIR) range. We now discuss these two observations in more detail.

A prominent asymmetric peak in $\sigma_1$ at 4.64 eV can be attributed to excitonic renormalization of the independent particle optical transitions at the M point (i.e. van Hove singularity (VHS)) in the Brillouin zone of the graphene band structure. If one considers only direct band to band transitions using a local density approximation (LDA) the optical transition peak should occur at $\omega \sim 4.1$ eV. By accounting for e-e interactions within a GW approach the optical transition peak is predicted to lie at 5.2 eV, and further incorporating e-h interactions the peak is red-shifted by ~600 meV from 5.2 eV to 4.6 eV [4-8]. This is precisely the position of the peak measured here for the pristine sample, as seen in Fig. 6.

In order to quantify the interplay between e-e and e-h interactions that manifests itself in the renormalization of the bare VHS peak in our controlled disorder graphene we employ the Fano phenomenological approach since, as seen previously, the asymmetric peak measured here at 4.64 eV resembles a Fano profile. Following Fano's model, the asymmetric line shape in the optical spectra can be thought of as arising from the coupling of the continuum electronic states near the saddle point singularity (M point) to discrete sharp excitonic states [24,25]. The resultant $\sigma_1$ in the presence of this Fano resonance can be described by [7]:

$$\frac{\sigma_1}{\sigma_{1,cont}} = \frac{(q+\varepsilon)^2}{1+\varepsilon^2}, \tag{5}$$

where $\sigma_{1,\text{cont}}(\omega)$ represents the continuum contribution to $\sigma_1$ arising from band-to-band transitions at the M point (possibly renormalized by e-e interactions [5]), $\varepsilon = \frac{2(\omega - E_{\text{res}})}{\Gamma}$, $\Gamma$ is a phenomenological width related to the lifetime of the exciton, and $E_{\text{res}}$ denotes the resonance energy. The parameter $q$ determines the lineshape, while $q^2$ gives the ratio of the strength of the excitonic transition to the unperturbed band-to-band transition. For simplicity, to capture quantitatively the spectral features in the vicinity of the resonance, we model $\sigma_{\text{cont}}$ with a constant background plus the logarithmic singularity in the joint density of states (JDOS) at the M point: $\sigma_{1,\,\text{cont}}(\omega) = -A\,log\left|1 - \omega/E_0\right| \otimes R_G(\omega) + C$, where A is a scaling factor, $E_0$ the unperturbed band to band transition energy, and $R_G(\omega)$ is a Gaussian of width $E_{\text{Br}} = 0.25$ eV that is convoluted with the bare divergence to account for the finite broadening [7]. The constant background (C) is intended to capture the universal optical conductivity ($\sigma_0$) of graphene at low frequencies.

Figure 7(a) shows the best fit of $\sigma_1$ for pristine single layer graphene using the phenomenological Fano analysis with the parameters $q = -1.1$, $\Gamma = 0.93$ eV, $E_{res} = 4.94$ eV, and $E_0 = 5.2$ eV. Despite the simplicity of the assumption used for $\sigma_{\text{cont}}$, the result of fitting eq. (5) to the data in the energy range between 0.5 to 5.0 eV captures the overall behaviour of the conductivity very well, including the universal $\sigma_1$ at lower energy and the main features of the renormalized van Hove peak. Figure 7(b) shows the separate contributions of $\sigma_{1,\text{cont}}$ (upper panel) and the Fano resonance (lower panel) to the resulting $\sigma_1$, as a function of the oxygen exposure time. Figures 8(a), (b), and (c) show the dependence of Fano parameters $q$, $\Gamma$, $E_{Res}$, and $E_0$, respectively, on the amount of disorder. The symmetric peak at 5.25 eV in $\sigma_{cont}$ coming from the unperturbed band-to-band transitions (see Fig.7(b)) decreases significantly to 50% in intensity without any shift in energy, as quantified in Figure 8(c). As for the contribution of the Fano resonance, it shows no significant change in shape as inferred from their $q$ and $\Gamma$ values, which barely change within the error bars (Figures 2(a) and (b)). However, a gradual blue shift as high as 100 meV is observed in the energy of the Fano resonance, as seen in Figure 3(c), where we show the position of the Fano resonance and the peak in $\sigma_{cont}$ that best fit each experimental curve for $\sigma_1(\omega)$. This variation of $E_{Res}$ suggests a considerable reduction of the excitonic binding energy (by 100 meV for $t_s$=6 s) upon oxygen exposure, whereas the persistence of the line shape indicates that, even though weakened upon oxygen exposure, the nature of the interaction itself has not changed. We consider now the dramatic suppression observed in $\sigma_1$ at frequencies below 1eV. Oxygen plasma hole-dopes graphene and might, or might-not, introduce strong renormalization of the band structure, depending on the amount of disorder, and how the oxidation affects the graphene lattice. Given that the overall profile of $\sigma_1(\omega)$ retains all the features of pristine graphene, we interpret the suppression of $\sigma_1$ at low frequencies as due to simple Pauli blocking, which excludes inter-band transitions for frequencies below $2E_F$ (see Figures 9(b) and 4(c)) [2]. For definiteness, and given that our lower experimental frequency limit is 0.5 eV, we explicitly compare the curve at the highest exposure time (that shows a clear suppression of absorption) with the expected frequency-dependent conductivity of doped SLG. Within a Kubo approach it reads [32]

$$\frac{\sigma_1(\omega)}{\sigma_0} = \frac{8}{\pi}\left(\frac{2\gamma k_B T}{\omega^2+4\gamma^2}\right) \log\left[2\cosh\left(\frac{E_F}{2k_B T}\right)\right] + f\left(-\frac{\omega}{2}\right) - f\left(\frac{\omega}{2}\right), \qquad (6)$$

where $2\gamma$ represents the half-width of the Drude peak and $f(E)$ is the usual Fermi-Dirac distribution function. The dashed line below $\omega$=1.4 eV in Fig. 6 shows the best fit of this expression to the experimental data at 300 K. From it we obtain $E_F$=0.558±0.003 eV and $\gamma$=0.021±0.004 eV ($\approx$170 cm$^{-1}$). Despite the absence of data below 0.5 eV for direct confirmation, this value is in reasonable agreement with the Drude width expected for graphene grown by CVD on Cu under high electron doping (>100 cm$^{-1}$) [33]. Assuming that the graphene dispersion remains linear after oxygen exposure, we can estimate the number of charge carriers ($n_e$) per unit cell as $n_e = A \frac{|E_F|^2}{\pi(\hbar v_F)^2}$, where $A$ is the area of the graphene unit cell, and $E_F$ can be extracted from the fit to the equation above. Figure 8(d) shows the dependence of $n_e$ on oxygen exposure. Moreover, in graphene the integrated spectral weight ($W$) in the NIR is conserved, and density independent: $\int_0^{\omega_M} \sigma(\omega)d\omega \simeq \sigma_0 \omega_M$, when the integration limit $\omega_M \gg 2E_F$. This integrated spectral weight is shown in Figure 9(a), and allows us to check the consistency of the extracted $E_F$ directly from the relative changes in optical spectral weight with different exposure times. The doping scenario is also consistent with the slight exposure-induced shift of the Raman G peak in our Raman data in Figure 1 [18,26]. The fact that the peak position changes only ~40 meV (Fig. 8(c)) from $t_s$= 2s to 6s while $n_e$ changes by an order of magnitude, (Fig. 8(d)) suggests the complex interplay between disorder and doping in the optical spectra.

What is striking in our optical data in this region is the very large suppression of $\sigma_1$ below $2E_F$ (Figure 6, $t_s$ = 6s), when the optical response of doped graphene in the Pauli-blocked region is usually characterized by a residual conductivity ~0.2-0.4$\sigma_0$ [27]. Such a residual conductivity can be justified theoretically on the basis of a finite electronic self-energy whose imaginary part is linear in $\omega$ [28]. The self-energy contributions can arise from the marginal Fermi liquid character of the electron-electron interactions in graphene, as well as optical phonons or disorder [29,30]. Our Raman data shows that the optical phonons are clearly affected by the oxygen exposure, and the Fano analysis of the optical data around the VHS reveals the clear suppression of the excitonic binding energy, thus hinting at reduced interactions with increased exposure times. Together these effects can lead to the suppression of the marginal Fermi liquid self-energy, thus explaining the nearly zero optical absorption below $2E_F$. Since our samples are disordered, it would also imply that the dominant mechanism for the residual optical conductivity in the Pauli-blocked region might indeed lie in interaction effects, rather than disorder. Finally, we underline that our plasma exposure is much milder than the intensities employed in recent reports, which reveal pristine graphene transitioning from ambipolar metallic to insulating behaviour upon treatment with oxygen plasma [18,19], whereas our samples retain the overall graphene signature in the frequency dependent optical conductivity.

## IV. Conclusion

In summary, we reported ellipsometry and Raman spectroscopic measurements on monolayer graphene exposed to *mild* oxygen plasma. We see that it affects the magnitude of electronic interactions from the reduction of the excitonic binding energy in the UV range, and from the nearly zero residual optical conductivity in the Pauli-blocked NIR range, which, according to existing interpretations for that residual optical absorption, is consistent with the weakening of the electronic self-energy. Our data suggests, in addition, that low levels of oxygen plasma exposure can be used as a controllable means to tune the optical absorption in graphene, without disrupting the overall graphene optical response, contrary to what happens in other oxygenation or doping strategies. Finally, it remains an open question with regard to IR spectral region below 0.5 eV where important details might lie for the full picture of the effects of mild oxygenation, including details of the Drude region and spectral weight redistribution. The results reported here certainly warrant further investigation of this spectral region in the future.

**Acknowledgment**

We are grateful to Antonio H. Castro Neto for very useful discussions. This work is supported by Singapore National Research Foundation under its Competitive Research Funding scheme of "Control of Exotic Quantum Phenomena at Strategic Interfaces and Surfaces for Novel Functionality by in-situ Synchrotron Radiation" and "Tailoring Oxide Electronics by Atomic Control", and MOE-AcRF-Tier-2, NUS-YIA and FRC. VMP is supported by the NRF-CRP grant R-144-000-295-281.

**FIGURES AND CAPTIONS:**

**Figure 1:** Raman spectra (excitation wavelength of 514 nm) for pristine single layer graphene (SLG) on SiO$_2$ (300 nm)/Si, and after consecutive intervals of oxygen plasma exposure: $t_s$ = 2s, 4s, and 6s.

**Figure 2:** Spectroscopic ellipsometry data of single layer chemical vapor deposition (CVD) graphene on SiO$_2$ (300 nm)/Si substrate. **(a)**. Ψ and **(b)** Δ for substrate (CVD graphene) shown in red (blue). The inset in (a) shows the optical multilayer model used in extracting Ψ and Δ. Media 0,1,2,3 in the model denote air, a graphene layer with the thickness $d_1$ of 0.335 nm, a SiO$_2$ layer with the thickness $d_2$ of 300 nm, and silicon bulk, respectively.

**Figure 3:** The experimental **(a)** Ψ and **(b)** Δ at an incident angle (θ$_0$) of 70° for pristine graphene on SiO$_2$ (300 nm)/Si substrate and after cumulative oxygen plasma exposure ($t_s$) in 2 seconds steps. The experimental **(c)** Ψ and **(d)** Δ at an incident angle (θ$_0$) of 70° for SiO$_2$ (300 nm)/Si substrate only. The insets amplify Ψ and Δ at particular energy ranges.

**Figure 4:** Detail of the analysis of spectroscopic ellipsometry data for pristine graphene on the SiO$_2$ (300 nm)/Si substrate. The experimental **(a)** Ψ and **(b)** Δ at different incident angles (θ$_0$)70°,65°, and 55° are shown in the solid blue, green, and red lines, respectively. The best match model extracted from the analysis is shown in dashed black lines. **(c)** Model dielectric constant used in the optical model for 300 nm SiO$_2$ layer. **(d)** Model dielectric constant used in the optical model for silicon bulk. **(e)** Model dielectric constant extracted from the optical model for the 0.335 nm graphene layer. **(f)** Corresponding optical conductivity of the graphene extracted from (e).

**Figure 5:** Detail of the analysis of spectroscopic ellipsometry data for graphene on the SiO$_2$ (300 nm)/Si substrate after 6s of oxygen plasma exposure time ($t_s$). The experimental **(a)** Ψ and **(b)** Δ at different incident angles (θ$_0$) 70° and 60° are shown in the solid blue and green lines, respectively. The best match model extracted from the analysis is shown in dashed black lines.

(**c**) Model dielectric constant used in the optical model for the 300 nm $SiO_2$ layer. (**d**) Model dielectric constant used in the optical model for the silicon bulk. (**e**) Model dielectric constant $\varepsilon_1$ extracted from the optical model for 0.335 nm graphene layer. (**f**) Corresponding optical conductivity $\sigma_1$ of 0.335 nm graphene extracted from (**e**).

**Figure 6:** Corresponding real part of optical conductivity, $\sigma_1(\omega)$, extracted from spectroscopic ellipsometry measurements (see Supplementary Information for details). The dashed line below 1.4 eV is the best fit to the data for $t_s = 6s$, as discussed in the text. The inset shows this fit in more detail.

**Figure 7:** (**a**) Fit of the experimental conductivity using the phenomenological Fano line shape analysis discussed in the text for pristine single layer graphene on $SiO_2$ (300nm)/Si. Upper panel: the optical conductivity $\sigma_1$ extracted from spectroscopic ellipsometry is shown as red circles, and the best fit to eq. (1) by the dashed line; the unperturbed band to band component, $\sigma_{cont}$, is shown in green/solid. Lower panel: Isolated Fano contribution $\sigma_{Res.Fit}/\sigma_{cont}$. (**b**) Comparison of the two individual fitting components (unperturbed band to band transitions, upper panel, and Fano resonance profile, lower panel) as the oxygen plasma exposure time increased. The grey area denotes the low energy region below $\sim 2E_F$ where $\sigma_1$ is Pauli-suppressed, which is not captured by the Fano approach.

**Figure 8:** Fano parameters extracted from the fit of $\sigma_1$ as function of oxygen plasma exposure time $t_s$. The corresponding disorder parameter $L_a$ (the in-plane crystalline grain size) derived from $t_s$ is depicted on top of the graphs. (**a**) The Fano line shape parameter $q$. (**b**) The exciton lifetime within Fano's model. (**c**) The peak position of $\sigma_{cont}$ ($E_0$), and Fano resonance energy ($E_{res}$). (**d**) The number of charge carriers ($n_e$) extracted directly from the Pauli blocking seen in the experimental traces of $\sigma_1(\omega)$.

**Figure 9:** (**a**) The integrated optical spectral weight ($W$) up to the photon energies in the horizontal axis. (**b**) Allowed optical transitions (vertical red arrow) in pristine graphene. $E_d$ and $E_F$ denotes the Dirac point and Fermi energy, respectively. (**c**) Likewise, for hole-doped

graphene. Optical transitions below $E_s$ are disallowed due to Pauli's exclusion principle. (**d**) In hole-doped and gapped graphene.

**FIGURE 1:**

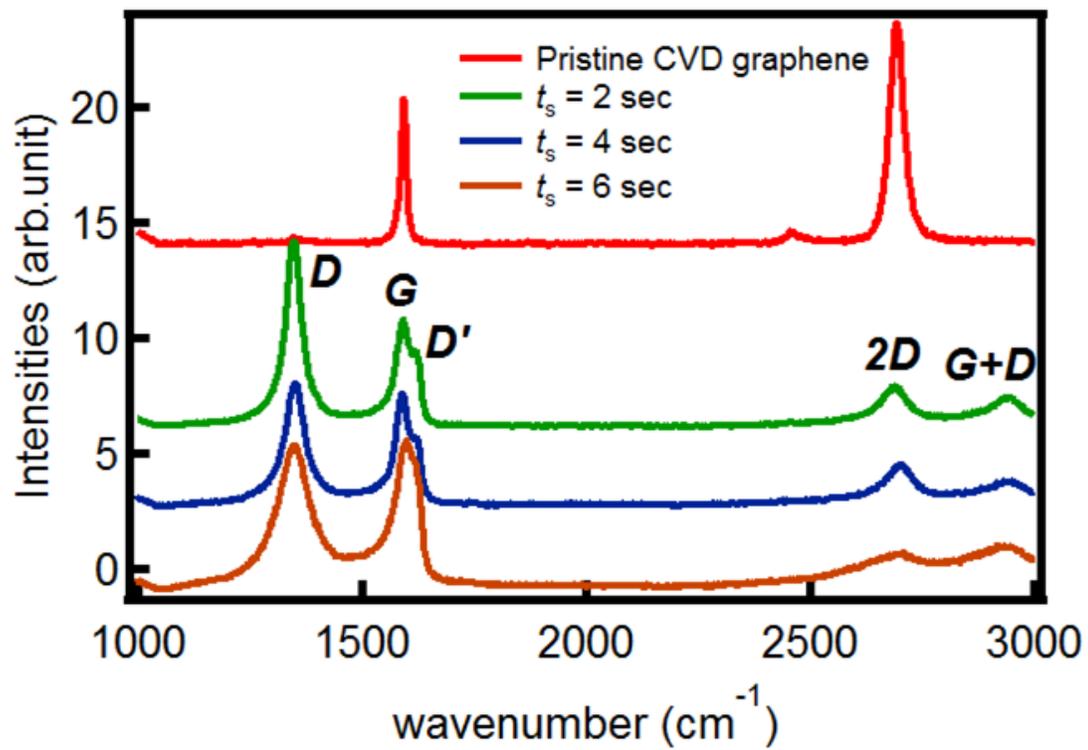

**FIGURE 2:**

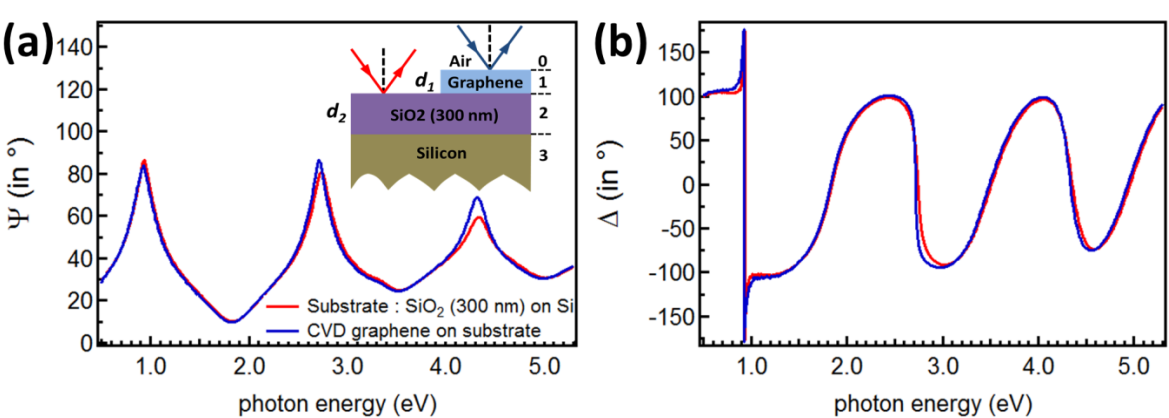

**FIGURE 3:**

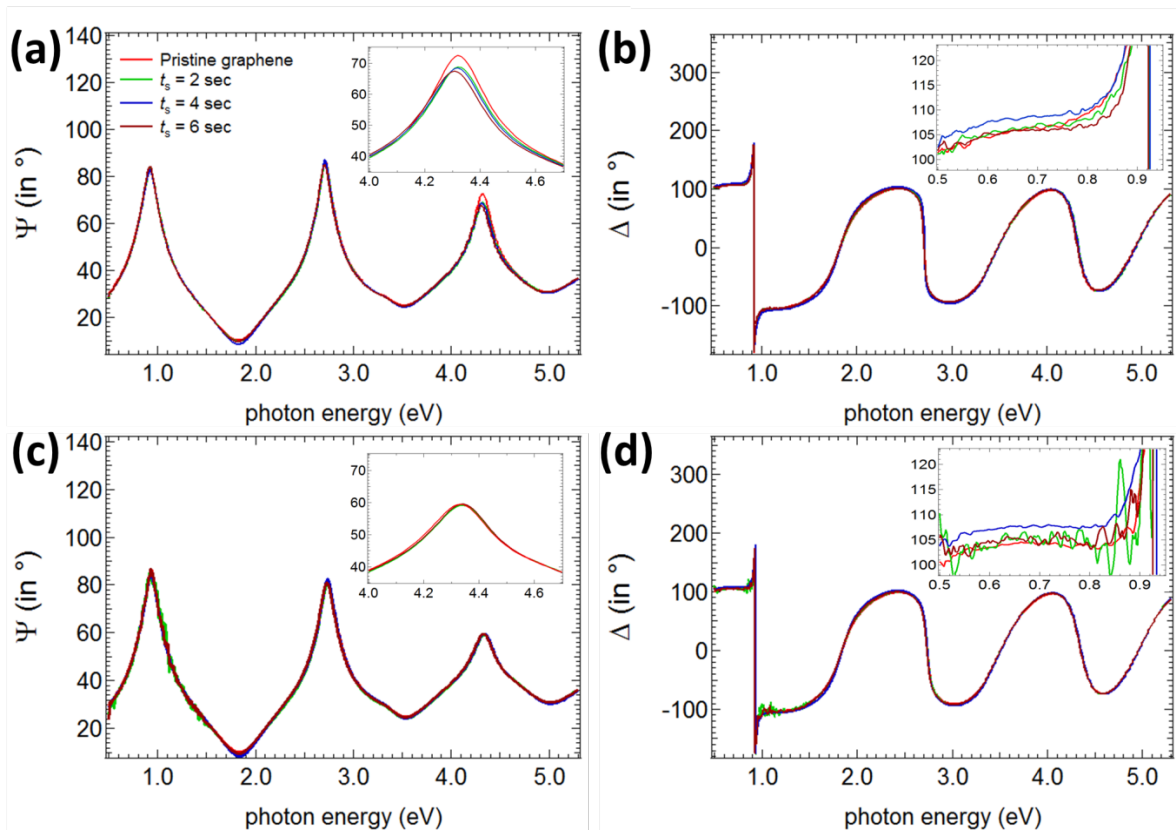

**FIGURE 4:**

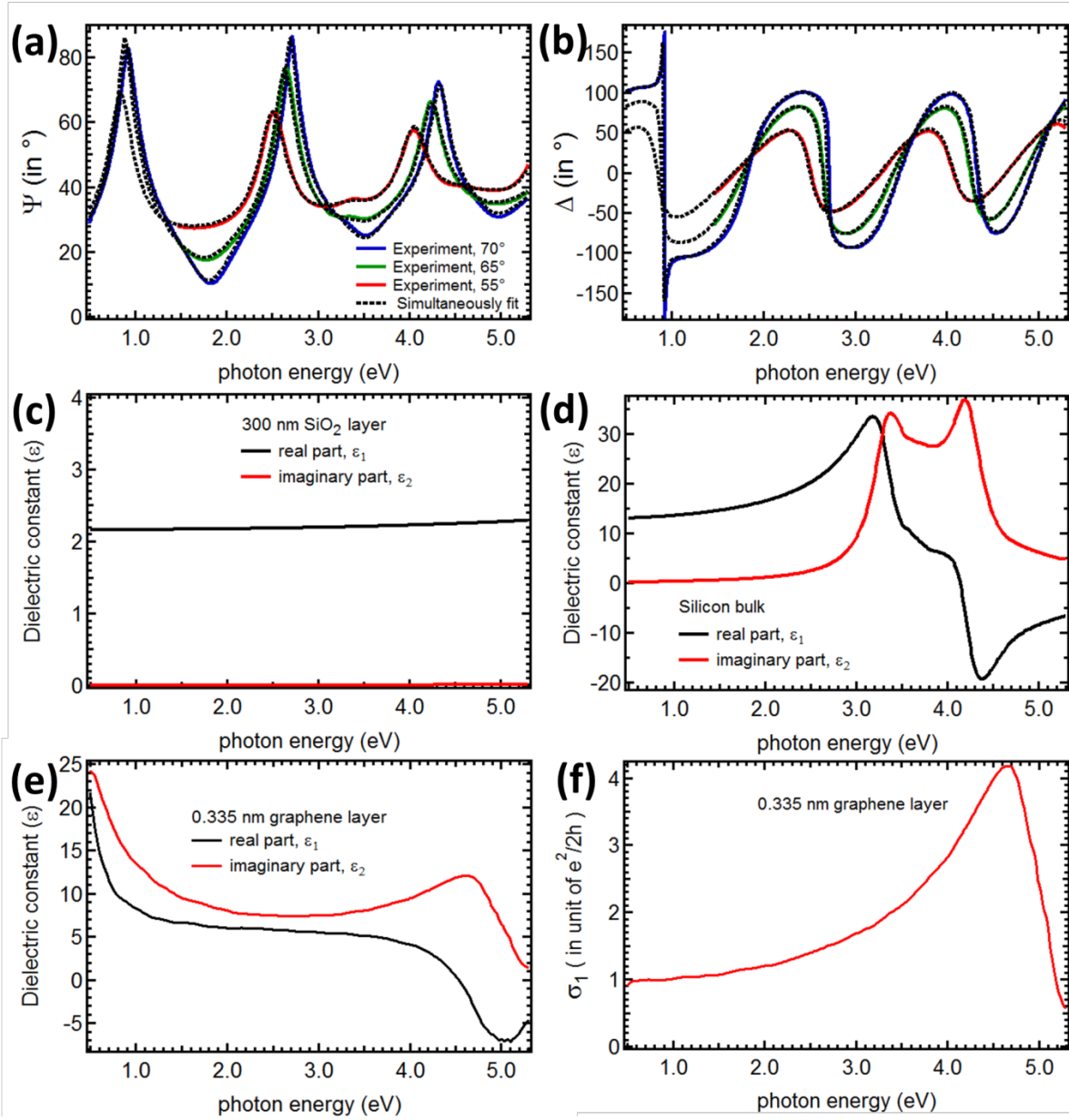

**FIGURE 5:**

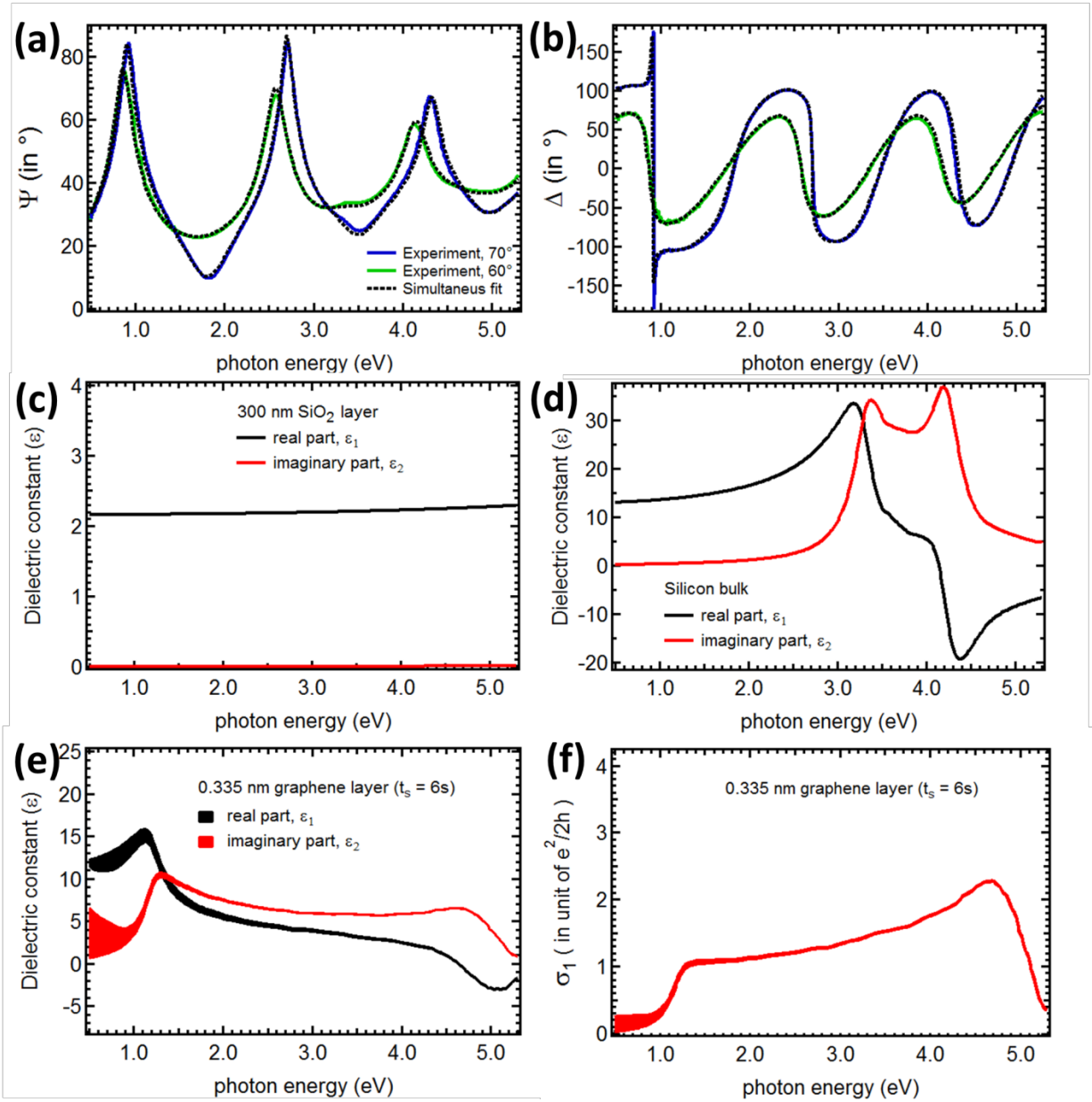

**FIGURE 6:**

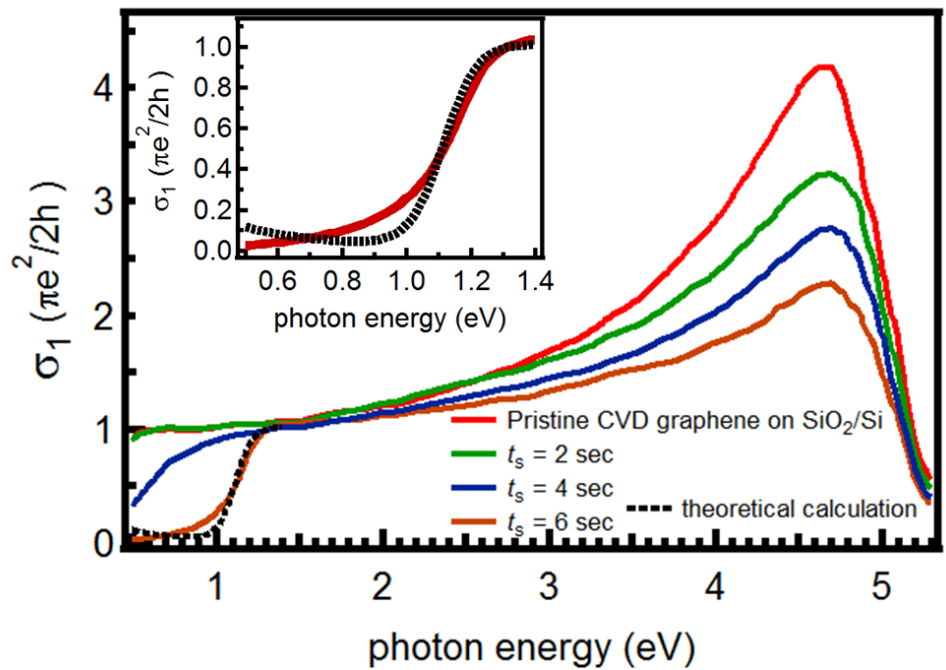

**FIGURE 7:**

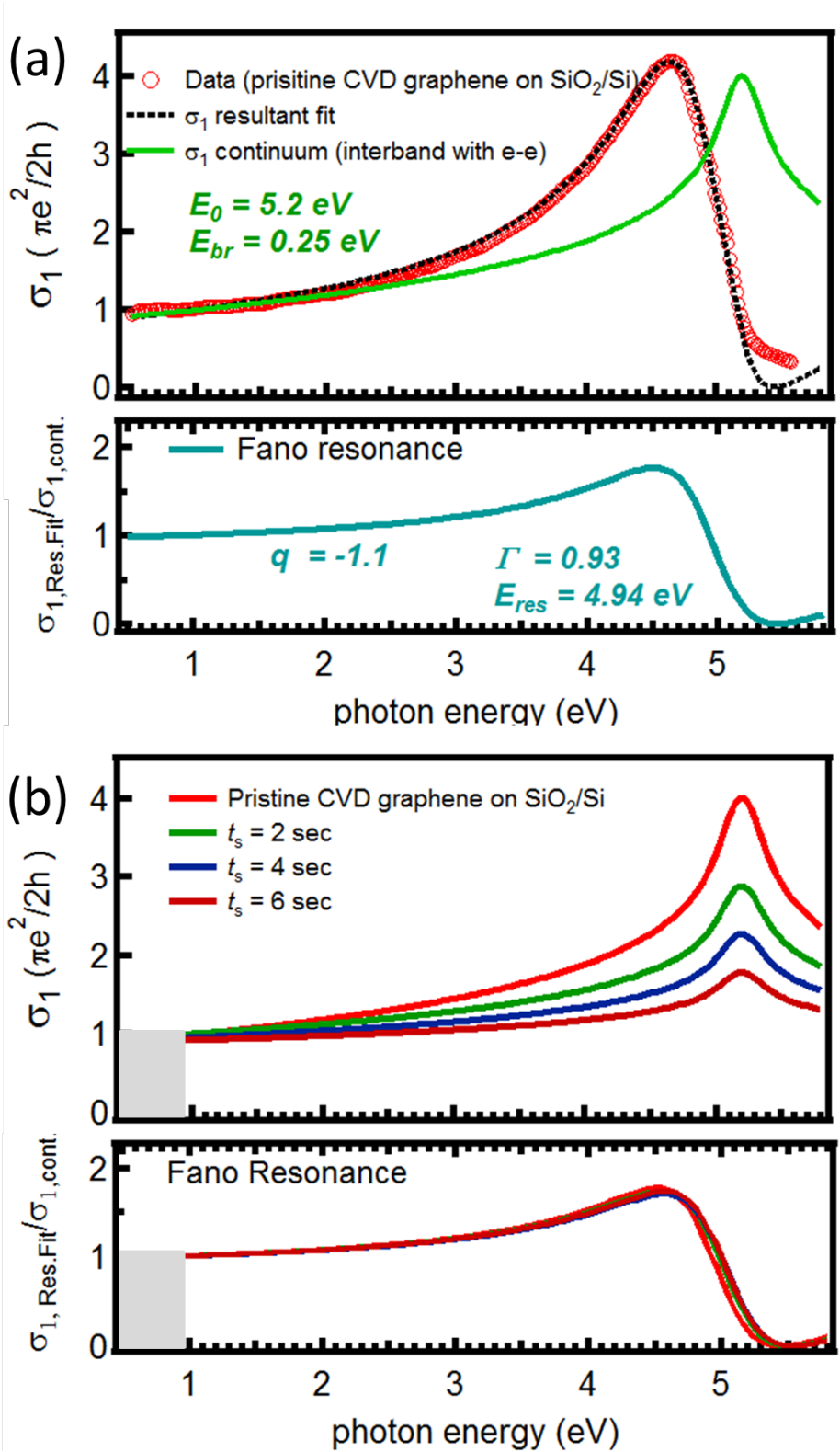

**FIGURE 8:**

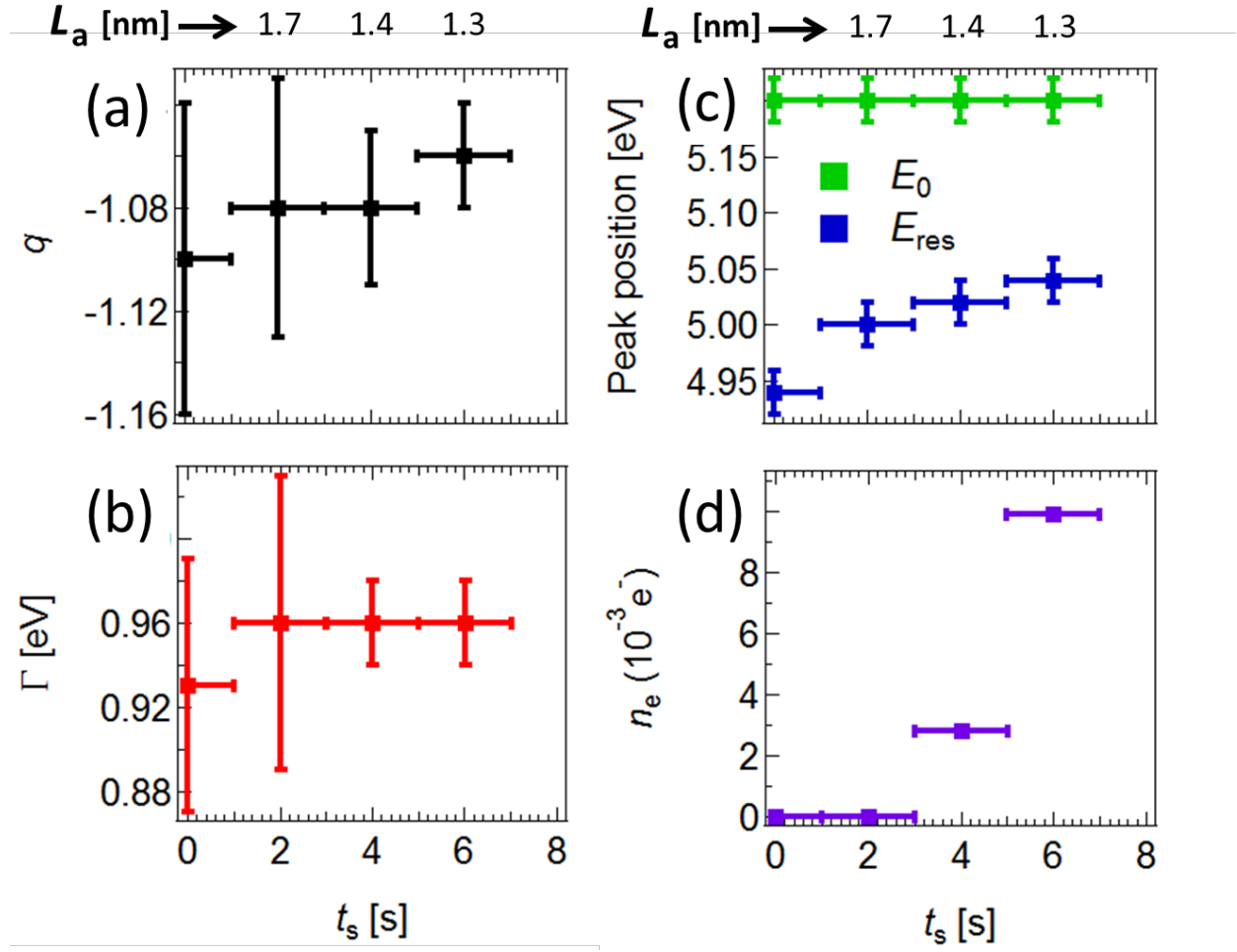

**FIGURE 9:**

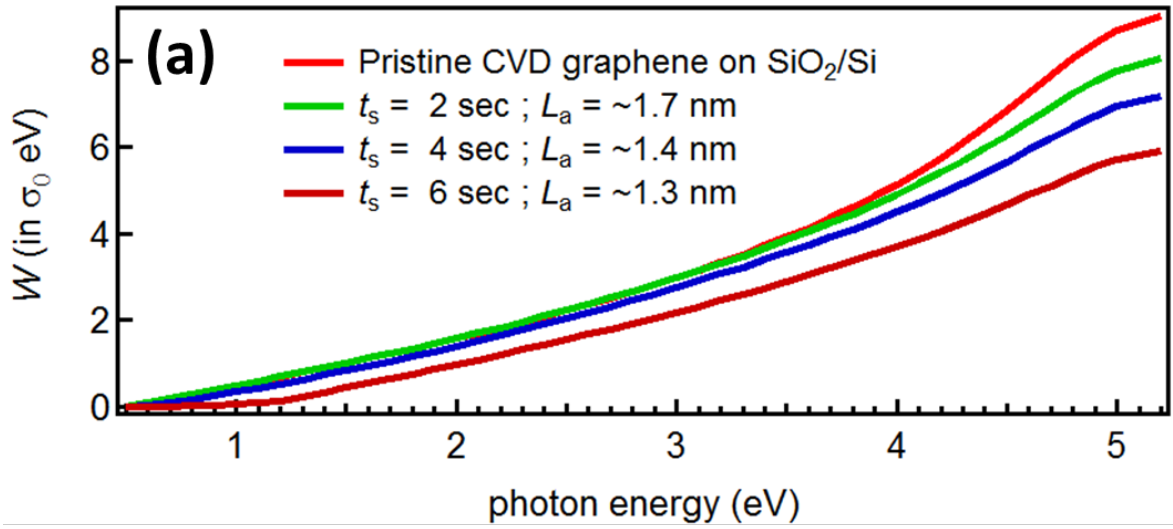

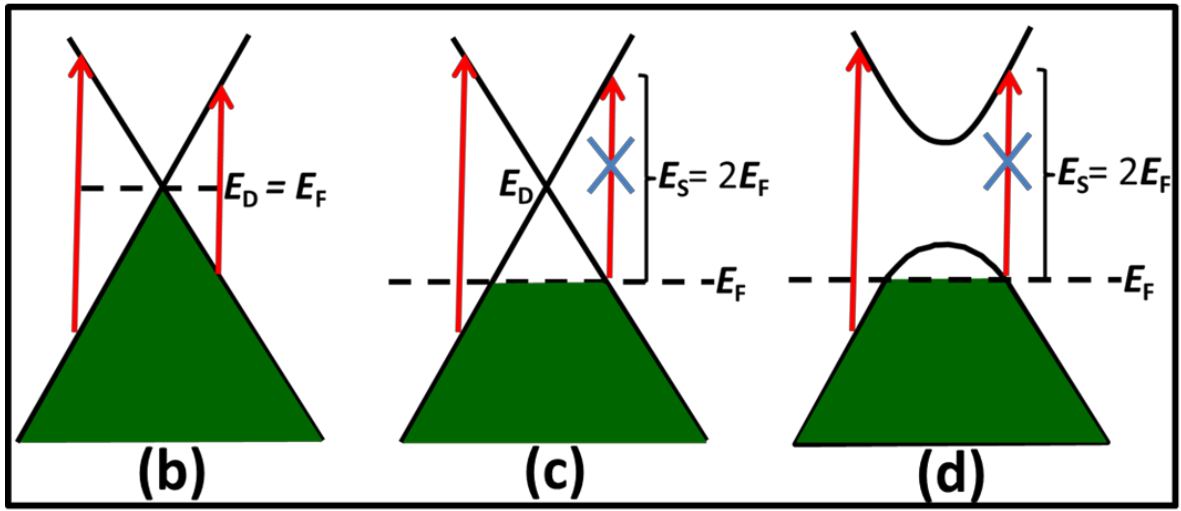